\documentclass[twocolumn,english]{revtex4}
\usepackage[T1]{fontenc}
\usepackage[latin9]{inputenc}
\usepackage{amsmath}
\usepackage{graphicx}
\usepackage{esint}

\makeatletter
\@ifundefined{textcolor}{}
{%
 \definecolor{BLACK}{gray}{0}
 \definecolor{WHITE}{gray}{1}
 \definecolor{RED}{rgb}{1,0,0}
 \definecolor{GREEN}{rgb}{0,1,0}
 \definecolor{BLUE}{rgb}{0,0,1}
 \definecolor{CYAN}{cmyk}{1,0,0,0}
 \definecolor{MAGENTA}{cmyk}{0,1,0,0}
 \definecolor{YELLOW}{cmyk}{0,0,1,0}
}

\makeatother

\usepackage{babel}
\begin{document}

\title{Competition of Lattice and Basis for Alignment of Nematic Liquid
Crystals}

\author{Andrew DeBenedictis}

\affiliation{Department of Physics and Astronomy, Tufts University, 574 Boston
Avenue, Medford, MA 02155, USA}

\author{Candy Anquetil-Deck}

\affiliation{Materials and Engineering Research Institute, Sheffield Hallam University,
City Campus, Howard Street, Sheffield, S1 1WB, UK}

\author{Douglas J. Cleaver}

\affiliation{Materials and Engineering Research Institute, Sheffield Hallam University,
City Campus, Howard Street, Sheffield, S1 1WB, UK}

\author{David B. Emerson}

\affiliation{Department of Mathematics, Tufts University, 503 Boston Avenue, Medford,
MA 02155, USA}

\author{Mathew Wolak}

\affiliation{Department of Mathematics, Tufts University, 503 Boston Avenue, Medford,
MA 02155, USA}

\author{James H. Adler}

\affiliation{Department of Mathematics, Tufts University, 503 Boston Avenue, Medford,
MA 02155, USA}

\author{Timothy J Atherton}

\email{timothy.atherton@tufts.edu}

\affiliation{Department of Physics and Astronomy, Tufts University, 574 Boston
Avenue, Medford, MA 02155, USA}
\begin{abstract}
Due to elastic anisotropy, two-dimensional patterning of substrates
can promote weak azimuthal alignment of adjacent nematic liquid crystals.
Here, we consider how such alignment can be achieved using a periodic
square lattice of circular or elliptical motifs. In particular, we
examine ways in which the lattice and motif can compete to favor differing
orientations. Using Monte Carlo simulation and continuum elasticity
we find, for circular motifs, an orientational transition depending
on the coverage fraction. If the circles are generalised to ellipses,
arbitrary control of the effective alignment direction and anchoring
energy becomes achievable by appropriate tuning of the ellipse motif
relative to the periodic lattice patterning. This has possible applications
in both monostable and bi-stable liquid crystal device contexts. 
\end{abstract}
\maketitle

\section{Introduction}

Surface anchoring, the promotion of a desired liquid crystal (LC)
orientation at a surface \citep{Jerome1991}, remains an important
problem in applications because precise tuning of anchoring parameters
is often necessary for optimal device performance \citep{spencer2006}.
Patterning the surface with a spatially varying preferred orientation
is an attractive route to create alignment layers with desired anchoring
properties because both the \emph{effective} anchoring strength and
its orientation or \emph{easy axis} can be altered by adjusting geometric
features of the pattern \citep{kondrat2003,barbero1992}. Additionally,
surfaces of appropriate symmetry \citep{Anquetil-Deck2012} may promote
multiple stable easy axes leading to \emph{bistable} devices \citep{Kim2001,Yi2008,Yoneya2002a,Lee2001}.
Bistability is desirable \citep{Kim2002,Stalder2003,Bryan-brown2000,Kitson2002}
both for reduced power consumption and improved addressing of hi-resolution
displays. Beyond displays, patterned LC systems are promising candidates
as biosensors \citep{Hwang2006,Lowe2010} and photonic devices \citep{Ruan2003,Wei2009}. 

Many methods exist to achieve patterning, encompassing both topographical
and chemical approaches. These include mechanical rubbing \citep{Bechtold2005,Varghese2004},
photolithography \citep{Bechtold2005,Schadt1992,Stalder2003}, scribing
with an atomic force microscope (AFM) \citep{Kim2002,Lee2004}, microcontact
printing of self-assembled monolayers (SAMs) \citep{Gupta1997,Cheng2000,Prompinit2010,Zheng2013},
and topographic surface features \citep{Liu2012}. Since mechanical
methods, such as rubbing, result in unwanted scratches or debris on
the surface \citep{Park2008}, and many methods do not scale well
to high-volume manufacturing \citep{Zheng2013}, SAMs have received
much attention in recent years. Certain experimental methods show
control over the azimuthal director angle as well as the polar anchoring
angle \citep{Bechtold2005,Wilderbeek2003} - this is the focus of
the work presented in this paper.

Striped surfaces, incorporating alternating regions preferring planar
and vertical alignment, have been well studied \citep{Atherton2006,Atherton2009,Atherton2010,Ledney2011,Kondrat2005,Yi2013}.
For this pattern, the polar angle of the bulk LC is controlled by
the average polar easy axis on the surface; the azimuthal alignment
has an energy minimum aligned parallel or perpendicular to the stripe
orientation, depending on the ratio of the elastic constants \citep{Atherton2006}.
Square checkerboard lattices are more complicated: an anchoring transition
occurs in which the liquid crystal aligns with the lattice vectors
for relatively strong surface anchoring, but switches to the diagonal
for very weak anchoring \citep{Anquetil-Deck2012}. Finally, both
polar and azimuthal control over the bulk LC director orientation
may be achieved with a rectangular checkerboard lattice. In this arrangement,
certain rectangle ratios and anchoring strengths combine to shift
the preferred director azimuthal angle from alignment with a rectangle
edge to alignment diagonally across the rectangle \citep{Anquetil-Deck2013}.

For substrates constructed from squares or rectangles, the pattern
element determines the symmetry and periodicity of the patterning.
It is, however, straightforward to break this coupling by resorting
to non-space-filling pattern basis motifs, such as circles or ellipses,
and arranging these on a periodic lattice. Importantly, this approach
provides a systematic approach by which to introduce the additional
parameters needed to achieve truly independent control of polar and
azimuthal anchoring angles and, so, set these at arbitrary target
values.

In this paper, we use a continuum approach to determine the ground
state of a liquid crystal in contact with a surface formed of a square
lattice of circular or elliptical motifs. From this, we identify scenarios
for which the basis symmetry and the lattice symmetry may favor different
alignments. This yields bistable configurations which are free from
the constraints that pertain for square or rectangular motifs. The
paper is organized as follows: Monte Carlo simulation results with
circle patterns are presented in section \ref{sec:Simulations}. In
section \ref{sec:Continuum-Model}, an analytical continuum model
is constructed for this arrangement with the simplifying assumption
that the director lies at a constant azimuthal angle. We also construct
a numerical model that relaxes this assumption. Brief conclusions
are presented in section \ref{sec:Conclusions}.

\section{Simulations\label{sec:Simulations}}

\begin{figure}
\includegraphics{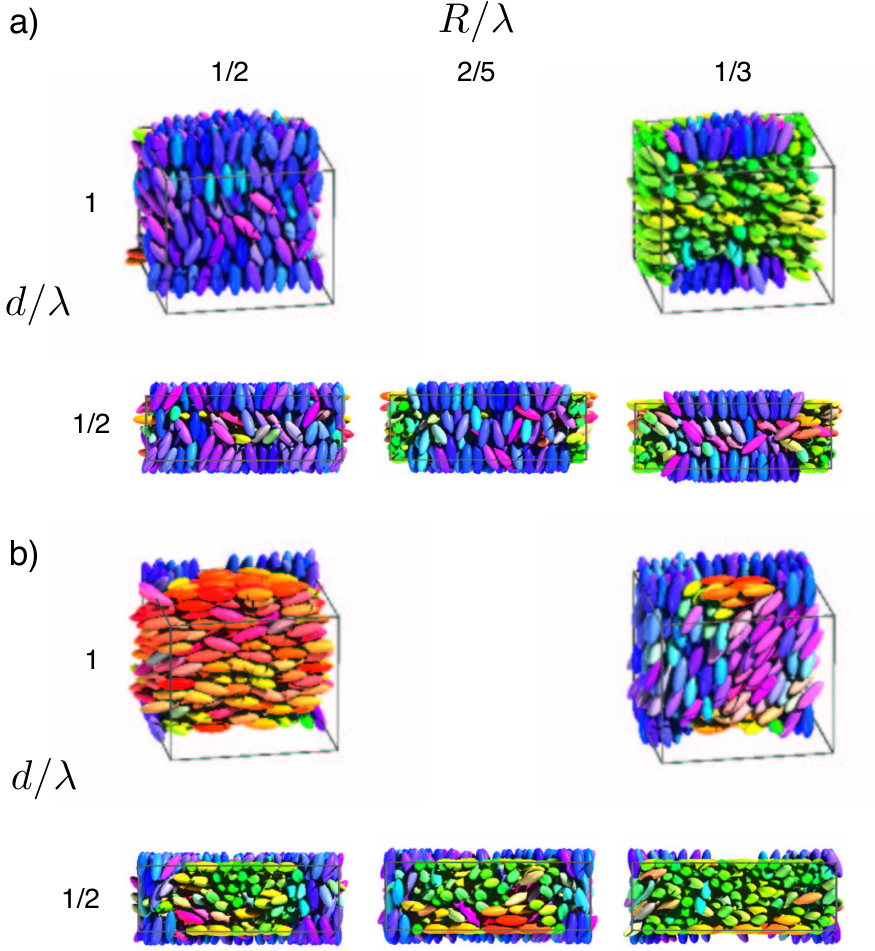}

\protect\caption{\label{fig:Snapshots}Snapshots of Monte Carlo simulations of a nematic
confined between periodically patterned circles. (a) Vertical circles
on planar background for different values of thickness $d$ and circle
radius $R$. (b) Planar circles on vertical background for $d/\lambda=1/2$;
the views show slices taken through the box mid-plane.}
\end{figure}

The combination of Monte Carlo (MC) simulations and continuum theory
has proven synergistic in previous studies \citep{Anquetil-Deck2012,Anquetil-Deck2013}
because the alignment induced by a particular pattern depends dramatically
on the length scales present: MC probes alignment around patterns
at the order of a few molecular lengths while continuum theory permits
modeling up to device scales.

To gain an initial understanding of the effect of circle patterns
on the adjacent nematic, we therefore first performed Monte Carlo
simulations as described fully in \citep{Anquetil-Deck2012}. Briefly,
particle-particle interactions are modeled with the hard Gaussian
overlap potential (HGO) and particle-wall interactions are modeled
with the hard needle-wall (HNW) potential, where a hard axial needle
of length $\sigma_{0}k_{s}$ is placed at the centers of particles
of short diameter $\sigma_{0}$. By adjusting the needle length, both
vertical ($k_{s}\le1)$ and planar alignment ($k_{s}>2$) can be achieved.

Here, simulations of HGO particles of aspect ratio $\kappa=3$ were
performed subject to confinement between two circle-patterned substrates
separated by a distance $d$. Periodic boundaries were imposed in
the $x$ and $y$ directions. Sharp boundaries were imposed between
vertical and planar regions, needle lengths being specified in each
region as described above. Particle configurations were initialized
at low density and uniformly compressed in the $x$ and $y$ until
orientational order was established. The corresponding in-plane box-length,
$\lambda$, then set the effective lattice periodicity of the chosen
circular motif. An equilibration run of $N=500,000$ MC sweeps was
conducted at each density, followed by a production run of a further
$N$ sweeps.

Representative snapshots are displayed in Fig. \ref{fig:Snapshots}(a)
for vertical circles on a planar background; corresponding plots for
the reverse case are in Fig. \ref{fig:Snapshots}(b). For $d=\lambda=4\kappa\sigma_{0}$,
the orientational ordering of the particles at the center of the film
can be seen to depend on $R$, the radius of the circle. If $d\simeq\lambda$,
an anchoring transition occurs with decreasing $R$. For large $R$,
the film follows the alignment of the particles in the circle be that
planar or vertical. However, a transition occurs with decreasing $R$,
after which the orientation in the film becomes dictated by the pattern
outside the circle. This behavior is observed for both vertical-on-planar
and planar-on-vertical surfaces and essentially mimics that seen for
other patterned films - the film orientation is dominated by the majority
surface component.

For thinner cells, where $d=\lambda/2$; the nematic follows the pattern
throughout the vertical distance for all values of $R$ studied. This
is similar to the ``bridging'' behavior observed in rectangular
patterned systems \citep{Anquetil-Deck2012,Anquetil-Deck2013}. For
the planar-on-vertical case {[}Fig. \ref{fig:Snapshots}(b){]}, the
particle orientation on the top and bottom substrates at a vertical-planar
boundary is particularly interesting, because the here the particles
tend to align parallel to the boundary causing azimuthal distortion
of the liquid crystal.

\begin{figure}
\includegraphics{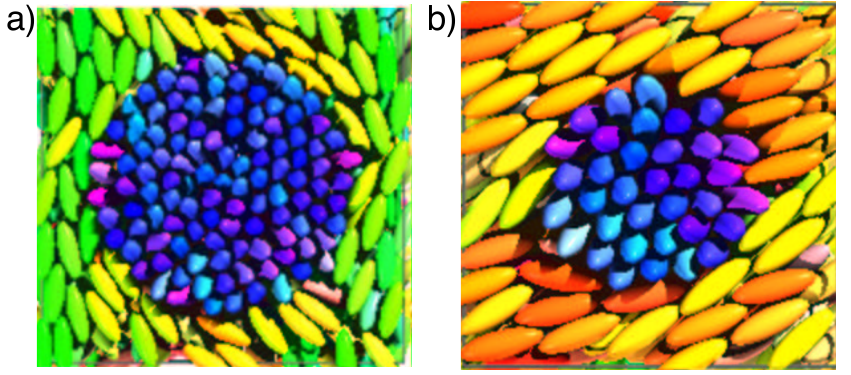}

\protect\caption{\label{fig:TopAndBottom}Apparent orientational anchoring transition.
a) Top configuration for vertical on planar circles with $R/\lambda=1/3$
and $d/\lambda=1$. b) Top configuration for $R/\lambda=1/4$.}
\end{figure}

A more subtle azimuthal transition is also apparent, however, from
careful study of vertical on planar systems. As illustrated in Fig.
\ref{fig:TopAndBottom}, the preferred azimuth of the planar region
is also dependent on $R$. Specifically, for larger circles particles
in the in-plane region aligned along the box $x$ and $y$-axes, whereas
at smaller $R$ they picked out the box diagonal. It is this azimuthal
transition that we focus on in the remainder of this paper.

\noindent 

\section{Continuum Model\label{sec:Continuum-Model}}

We now turn to the continuum analysis. The LC orientation is characterized
by a director field,

\begin{equation}
\mathbf{n}=(\cos\theta\sin\phi,\cos\theta\cos\phi,\sin\theta),\label{eq:director}
\end{equation}
where $\theta$ is the zenithal angle and $\phi$ is the azimuthal
angle. The free energy of the LC will be given by the Frank free energy

\begin{eqnarray}
F & = & \frac{1}{2}\int d^{3}\!x\,[K_{1}(\nabla\cdot\mathbf{n})^{2}+K_{2}(\mathbf{n}\cdot\nabla\times\mathbf{n})^{2}\nonumber \\
 &  & +K_{3}|\mathbf{n}\times\nabla\times\mathbf{n}|^{2}]+\underset{S}{\int}dS\,g(\theta-\theta_{e}),\label{eq:frank}
\end{eqnarray}
with a harmonic anchoring potential

\begin{equation}
g(\theta-\theta_{e})=\frac{W_{\theta}}{2}(\left.\theta\right|_{z=\pm d/2}-\theta_{e})^{2}.\label{eq:RPpotential}
\end{equation}

\begin{figure}
\includegraphics{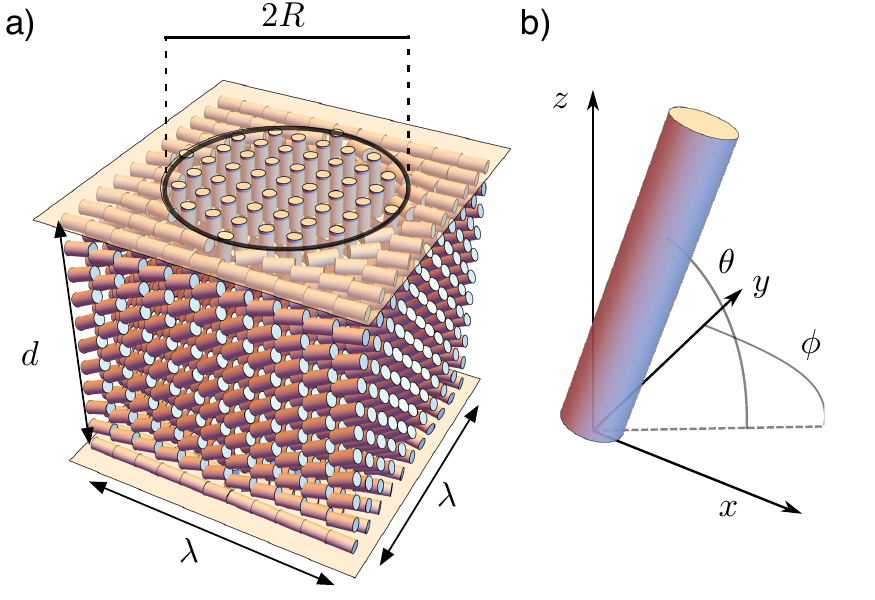}

\protect\caption{\label{fig:Coordinates}(a) Schematic of the unit cell domain with
important length scales labeled for a circle-patterned surface ($a=b=R$
and $\omega=0$). (b) Definition of the polar angle $\theta$ and
the azimuthal angle $\phi$ of the LC director.}
\end{figure}

The coordinates are set up as follows: consider a unit cell defined
on the box with corners at $(0,0,-d/2)$ to $(\lambda,\lambda,+d/2)$.
Each surface at $\pm d/2$ contains an ellipse centered on the unit
cell with semi-major axis $a$ oriented at an angle $\omega$ with
respect to the $x$-axis and semi-minor axis $b$. The surfaces promote
homeotropic ($\theta=\pi/2$) alignment within the ellipse and planar
alignment ($\theta=0$) outside, as shown in Fig. \ref{fig:Coordinates}. 

\subsection{Solution}

Following previous work, we make the two-constant approximation by
setting $K_{1}=K_{3}$ and $K_{2}/K_{1}=\tau$ \citep{Atherton2006}.
Additionally, if the polar angle depends on all of the coordinates
$\theta=\theta(x,y,z)$ and the azimuthal angle is constant in space
$\phi=\phi_{0}$ \citep{Anquetil-Deck2013}, the bulk free energy
may be rewritten as a quadratic form,

\begin{equation}
F=\frac{K_{1}}{2}\int\left(\nabla\theta\right)^{T}\cdot A\cdot(\nabla\theta)d^{3}x+\frac{K_{1}}{2L_{\theta}\lambda}\underset{S}{\int}dS\,(\left.\theta\right|_{z=\pm d/2}-\theta_{e})^{2}
\end{equation}
where

\[
A=\left(\begin{array}{ccc}
\frac{1}{2}(\alpha-\beta\cos2\phi_{0}) & \frac{1}{2}\beta\sin2\phi_{0} & 0\\
\frac{1}{2}\beta\sin2\phi_{0} & \frac{1}{2}(\alpha+\beta\cos2\phi_{0}) & 0\\
0 & 0 & 1
\end{array}\right)
\]
and we introduced the dimensionless polar anchoring parameter

\begin{equation}
L_{\theta}=\frac{K_{1}}{W_{\theta}\lambda}.\label{eq:anchorParam}
\end{equation}

Within this approximation, minimization of the free energy yields
an anisotropic Laplace equation,

\begin{equation}
\mathbf{\nabla}^{T}\cdot A\cdot\mathbf{\nabla}\theta(x,y,z)=0,\label{eq:laplace}
\end{equation}
where $\alpha\equiv1+\tau$ and $\beta\equiv1-\tau$ . This may be
solved using the series,

\begin{equation}
\theta(x,y,z)=\overset{\infty}{\!\!\underset{n,m=-\infty}{\sum}\!\!}2A_{nm}\cosh(r_{nm}z)e^{i2\pi(nx+my)},\label{eq:theta}
\end{equation}
where

\begin{equation}
\frac{r_{nm}}{\sqrt{2}\pi}=\sqrt{\alpha(m^{2}+n^{2})+\beta(m^{2}-n^{2})\cos2\phi_{0}+2\beta nm\sin2\phi_{0}}.\label{eq:rnm}
\end{equation}
In order to satisfy the boundary conditions, the patterned easy axis
is first expanded in a Fourier series,

\begin{equation}
\theta_{e}(x,y)=\overset{\infty}{\underset{n,m=0}{\sum}}S_{nm}e^{i2\pi(nx+my)}.\label{eq:easyaxis}
\end{equation}
To determine the coefficients $S_{nm}$, we assume that the background
surface promotes $\theta_{e}=0$ while the elliptical pattern promotes
$\theta_{e}=\pi/2$, henceforth referred to as \textquotedblleft vertical
on planar'' patterning; the alternative \textquotedblleft planar
on vertical\textquotedblright{} arrangement is trivially obtained
from this solution by making the substitution,
\begin{equation}
\theta\to\pi/2-\theta,\label{eq:transformation}
\end{equation}
which leaves the energy invariant. The $S_{nm}$ are therefore evaluated
by integrating,

\begin{equation}
S_{nm}=\frac{\pi}{2}\iint_{\mathcal{D}}\exp\big(2\pi i(nx+my)\big)dxdy,\label{eq:surfcoef}
\end{equation}
over a domain $\mathcal{D}$ defined by the ellipse equation,

\begin{equation}
\left(\vec{x}-\vec{x}_{c}\right)^{T}R(\omega)\left(\begin{array}{cc}
1/a^{2} & 0\\
0 & 1/b^{2}
\end{array}\right)R(-\omega)\left(\vec{x}-\vec{x}_{c}\right)\le1.\label{eq:domain}
\end{equation}
Here, $R(\omega)$ is the 2D rotation matrix and $\vec{x_{c}}=(\frac{1}{2},\frac{1}{2})$
is the center of the ellipse. As shown in the appendix, the integral
(\ref{eq:surfcoef}) can be performed analytically to yield,

\begin{equation}
S_{nm}=\frac{ab\pi}{2}(-1)^{n+m}\frac{J_{1}\left(2\pi\sqrt{a_{nm}^{'2}+b_{nm}^{'2}}\right)}{\sqrt{a_{nm}^{'2}+b_{nm}^{'2}}},\label{eq:surfcoefsol}
\end{equation}
where $J_{1}(x)$ is a Bessel function of the first kind, $a_{nm}^{'}=(n\cos\omega+m\sin\omega)a$,
and $b_{nm}^{'}=(n\sin\omega-m\cos\omega)b$. 

Having expanded the easy axis in a suitable form, the coefficients
$A_{nm}$ can be determined by imposing the Robin boundary condition
at $\pm d/2$.

\begin{equation}
\theta_{e}=\left[\pm L_{\theta}\frac{\partial\theta}{\partial z}+\theta\right]_{z=\pm d/2},\label{eq:robin}
\end{equation}
Here, the $\pm$ refers to the direction of the outward normal to
the LC boundary. Inserting Eqs. (\ref{eq:theta}) and (\ref{eq:easyaxis})
into (\ref{eq:robin}), we obtain,

\begin{equation}
A_{nm}=\frac{S_{nm}}{2\left(L_{\theta}r_{nm}\sinh\left(\frac{r_{nm}d}{2}\right)+\sinh\left(\frac{r_{nm}d}{2}\right)\right)}.\label{eq:coefsol}
\end{equation}
Note that as $L_{\theta}\to0$, this recovers the rigid anchoring
condition. The solution for $\theta(x,y,z)$ is now obtained by inserting
Eqs. (\ref{eq:coefsol}), (\ref{eq:surfcoefsol}), and (\ref{eq:rnm})
into Eq. (\ref{eq:theta}). 

As for other patterns \citep{Anquetil-Deck2012,Anquetil-Deck2013},
the director follows the surface pattern at $\pm d/2$, while relaxing
to a uniform orientation, equal to the average polar angle promoted
by the surface far away from the boundaries.

\subsection{Circular patterns}

For a circular surface pattern, set $a=b$ in Eq. (\ref{eq:surfcoefsol}).
Evaluation of the volume integral in Eq. (\ref{eq:frank}) then yields
an expression for the bulk energy of the LC,

\begin{widetext}

\begin{equation}
F_{b}=\overset{\infty}{\underset{n,m=-\infty}{\sum}}2K_{1}A_{nm}^{2}\left[B^{T}A\left(\frac{d}{2}\,\textnormal{diag}(1,1,-1)+\frac{\sinh\left(r_{nm}d\right)}{2r_{nm}}I_{3}\right)B\right]
\end{equation}
\end{widetext}where
\[
B=\left(\begin{array}{c}
2\pi n\\
2\pi m\\
r_{nm}
\end{array}\right).
\]
Similarly, evaluation of the surface integral over the surfaces at
$+d/2$ and $-d/2$ gives the surface energy of the LC,

\[
F_{s}=\overset{\infty}{\underset{n,m=-\infty}{\sum}}\frac{K_{1}\left(S_{nm}-2A_{nm}\cosh(\frac{r_{nm}d}{2})\right)^{2}}{L_{\theta}}.
\]

\begin{figure}
\includegraphics{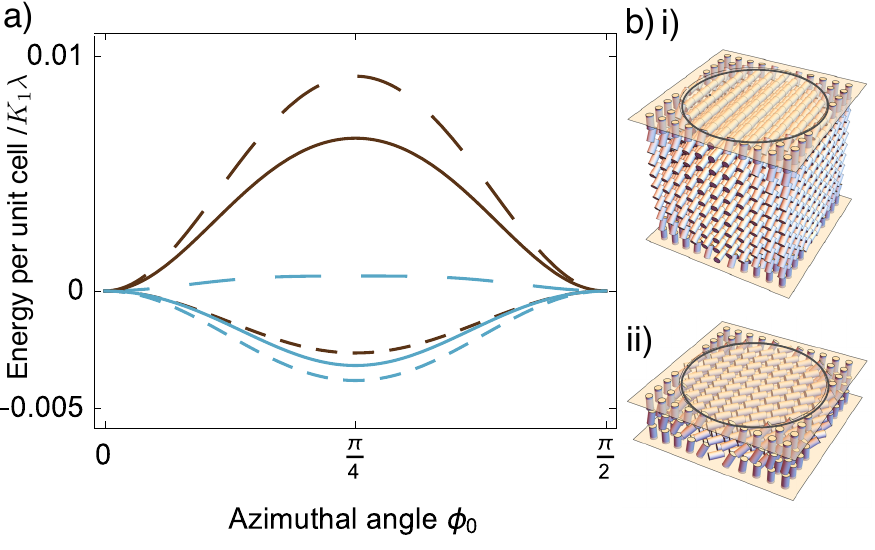}\protect\caption{\label{fig:eProfile}(a) LC surface energy (dotted), bulk energy (dashed),
and total energy (solid) as a function of the azimuthal angle for
a unit cell of thickness $d/\lambda=1.0$ (brown) and $d/\lambda=0.25$
(cyan) for planar on vertical patterning, $r/\lambda=0.5$, and $L=0.01$.
For the thicker unit cell, the surface energy azimuthal preference
is much weaker than that of the bulk and the overall preference aligns
with that of the bulk. However, thinner cells exhibit weaker bulk
energy preference and the total energy favors alignment along $\phi_{0}=\pi/4$.
(b) Corresponding calculated structures for i) $d/\lambda=1.0$ and
ii) $d/\lambda=0.25$. }
\end{figure}

The bulk and surface energy of the LC are shown in Fig. \ref{fig:eProfile}
as a function of the azimuthal angle $\phi$ for fixed circle radius
and polar anchoring energy. The bulk energy always prefers alignment
along the $x$- or $y$-axis, but with decreasing strength as the
unit cell thickness decreases. Meanwhile, the surface energy prefers
director alignment at a 45-degree angle to the axes. This surface
preference grows slightly stronger as cell thickness decreases. Results
for planar on vertical patterning are identical due to the invariance
of the energy under the linear transformation (\ref{eq:transformation}). 

\begin{flushleft}
\begin{figure}
\includegraphics{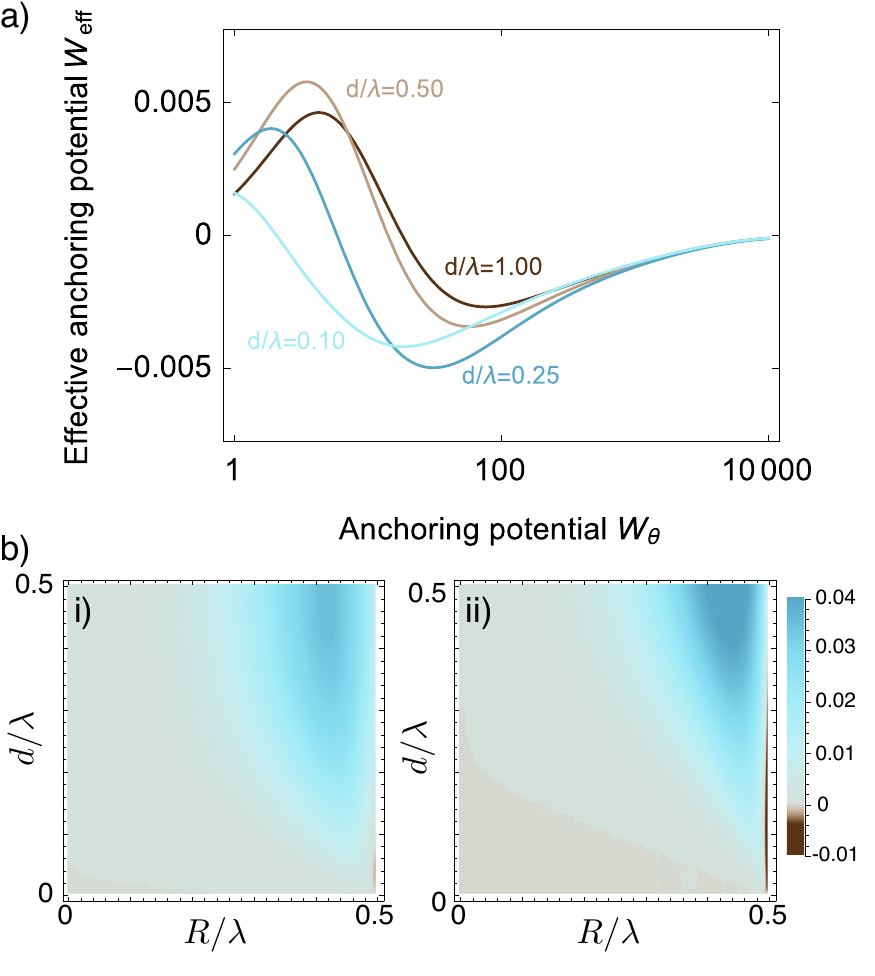}

\protect\caption{\label{fig:phaseDiagram}(a) Effective anchoring energy as a function
of polar anchoring energy $W_{\theta}$ for $R/\lambda=0.5$ and several
cell depths. As the cell depth decreases, the surface preference becomes
more pronounced for a given value of $W_{\theta}$. (b) Phase diagram
showing the strength and orientation of the preferred azimuthal alignment
angle as a function of cell depth $d/\lambda$ and circle radius $R/\lambda$
for i) $L_{\theta}=0.10$0, and ii) $L_{\theta}=0.001$. Brown regions
indicate diagonal alignment $\phi_{0}=\pi/4$ while cyan regions regions
prefer alignment with the lattice vectors, $\phi_{0}=0$ or $\phi_{0}=\pi/2$.
The brown regions expands into larger cell depths as the anchoring
strength increases. }
\end{figure}

\par\end{flushleft}

We estimate $W_{eff}$, the \textit{\emph{effective}}\textit{ }\textit{\emph{azimuthal
surface anchoring energy \citep{Harnau2007}}} as the energy difference
between the $\phi_{0}=\pi/4$ and the $\phi_{0}=0$ states per unit
area. A positive value of $W_{eff}$ indicates a preference for alignment
along the $x$- or $y$-axis, while a negative value of $W_{eff}$
indicates a preference for diagonal alignment. Fig. \ref{fig:phaseDiagram}(a)
shows the effective surface anchoring potential as a function of $W_{\theta}$,
the polar anchoring energy, for a series of cell thicknesses. The
surface contribution dominates for a thin cell with strong anchoring
$W_{\theta}\approx100$, which corresponds to $L_{\theta}=0.01$.
In the limit of rigid anchoring ($W_{\theta}\rightarrow\infty$) or
extremely weak anchoring ($W_{\theta}\rightarrow0$), the effective
anchoring vanishes. For weak anchoring, the nematic effectively ignores
the pattern, while for rigid anchoring the surface follows the prescribed
pattern exactly. 

To determine the parameter space in which parallel and diagonal alignment
are each favored, we display in Fig. \ref{fig:phaseDiagram}(b) the
strength of the azimuthal energy preference computed from the combined
bulk and surface energies, shown for $L_{\theta}=0.100$, and $L_{\theta}=0.001$.
The diagrams indicate two regions in which the the surface preference
overrides the bulk preference for a sufficiently thin cell. The strongest
of these regions is the thin brown section at $R/\lambda=0.5$, the
largest possible circle radius. This region of parameter space is
dominated by the surface term, likely due to the inability of vertically/horizontally
aligned LCs to effectively fill space between two abutting circles
whose edges are nearly perpendicular to the LC director. For circles
much smaller than $R=0.5$, the preference becomes very weak indeed.

The comparable magnitude and conflicting preferences of the bulk and
surface energies at these length scales suggests that the constant-$\phi$
approximation may be too restrictive for this system. Unlike the rectangular
and square patterns previously considered, the alignment direction
is promoted exclusively by the lattice while the basis favors no particular
alignment. The results of the Monte Carlo simulations also suggest
this: in the configurations shown in Fig. \ref{fig:Snapshots}(b),
the particles tend to align tangentially around the edge of the circle
because for $\tau<1$ the energetically cheapest way to achieve the
vertical-to-planar transition around the perimeter of the circle is
through a twist deformation. In section \ref{sub:Numerical-model},
therefore, we numerically minimize the free energy (\ref{eq:frank})
with respect to a completely arbitrary director profile to quantify
the effect of azimuthal variations.

\subsection{Elliptical patterns\label{sub:Ellipse-patterned-surfaces}}

We now consider elliptical patterns. For long ellipses, one might
expect the effective azimuthal alignment to lie parallel to the semi-major
axis, resembling the situation with alignment on striped surfaces\citep{Atherton2006}.
Hence, by adjusting $\phi$, it should be possible to control the
preferred azimuthal angle arbitrarily and, by tuning the aspect ratio,
also control the effective azimuthal anchoring energy. The control
parameter space to consider is greatly expanded: while the cell depth
$d/\lambda$ and the anchoring potential $W_{\theta}$ remain parameters,
the circle radius $R/\lambda$ is replaced by the semi major axis
length $b/\lambda$, aspect ratio $b/a$, and the alignment angle
$\omega$. From the structure of the solution (\ref{eq:theta}), we
expect the cell depth and anchoring potential to have similar effects
in both patterns and so we focus on the effects of the new parameters
in this section.

\begin{figure}
\includegraphics{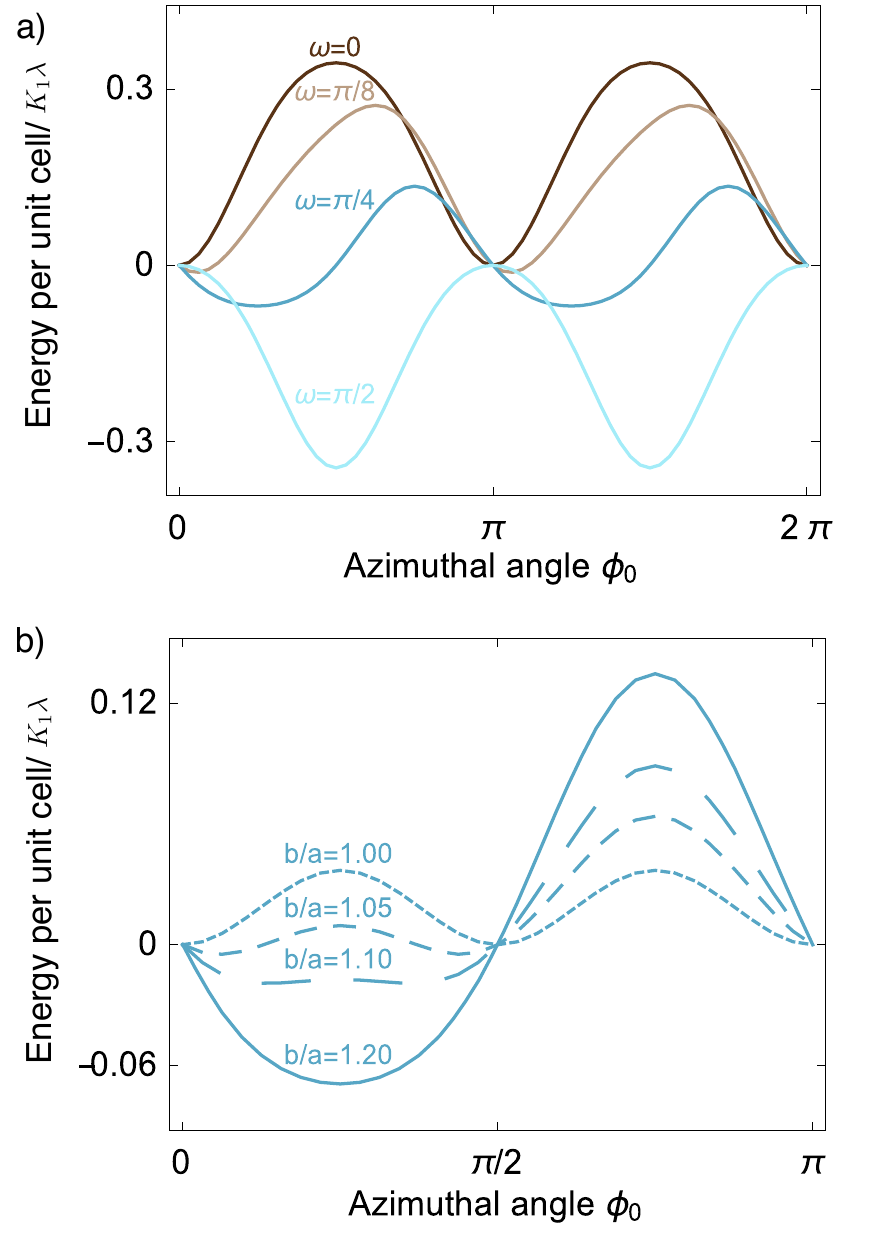}

\protect\caption{\label{fig:ellipseRotate}Total LC energy as a function of the azimuthal
angle shown for (a) several ellipse rotation angles with $b/a=1.2$
and (b) for several different aspect ratios with $\omega=\pi/4$.
Both panels have $d/\lambda=1$, $L_{\theta}=0.01$, and a coverage
fraction equal to that of a circle with $R/\lambda=0.4$. Rotation
of the ellipse results in migration of the preferred azimuthal angle
from alignment with the axes of the unit cell to alignment with the
semi major axis of the ellipse. However, panel (b) shows that this
transition in azimuthal angle preference is not immediate, but instead
passes through several smaller angles as aspect ratio increases.}
\end{figure}

In Fig. \ref{fig:ellipseRotate}, we show effective azimuthal anchoring
profiles $W_{eff}(\phi)$ calculated for a variety of values of $\omega$
and $b/a$. The coverage fraction of the pattern, or equivalently
the area of the elliptical motif, is kept constant in order to fix
the effective polar angle. An immediately obvious feature, compared
to the equivalent profiles for circular patterns plotted in Fig. \ref{fig:eProfile},
is that the mirror symmetry of the pattern about $\phi=\pi/2$ is
entirely broken, leaving behind a non-symmetric anchoring potential
reminiscent of the structures fabricated in \citep{Ferjani2010}.

For fairly modest aspect ratios, the alignment angle of the ellipse
controls the preferred azimuthal angle such that the energy minima
occur at $\phi_{0}=\omega$. For smaller values of $\omega$, or aspect
ratios close to unity, the results are more complex: For instance,
the energy minimum for $\omega=\pi/8$ in Fig. \ref{fig:ellipseRotate}(a)
occurs at $\phi_{0}=\pi/16$ instead of $\phi_{0}=\pi/8$. Also, in
Fig. \ref{fig:ellipseRotate}(b) we see that, for a rotation angle
of $\pi/4$, the azimuthal preference moves from alignment with the
sides of the unit cell to alignment with the semi major axis of the
ellipse gradually, preferring $\phi_{0}=\pi/16$ for an aspect ratio
of 1.05 and $\phi_{0}=\pi/8$ for an aspect ratio of 1.1.

The mechanism for this transition is two-fold. Firstly, the surface
energy consistently prefers azimuthal alignment along the semi-major
axis of the ellipse. Though the angle preferred by the surface remains
the same for any non-unit aspect ratio, the strength of the preference
grows with increasing aspect ratio. Second, as the aspect ratio grows,
the bulk energy of the LC tends to prefer an azimuthal angle aligned
with the ellipse alignment angle, but this move happens slowly such
that small-to-moderate aspect ratios result in a preferred angle somewhere
between $\phi_{0}=0$ and $\phi_{0}=\omega$. The magnitude of the
bulk energy preference also grows with increasing aspect ratio, but
less dramatically than that of the surface energy preference. 

\begin{figure*}
\includegraphics{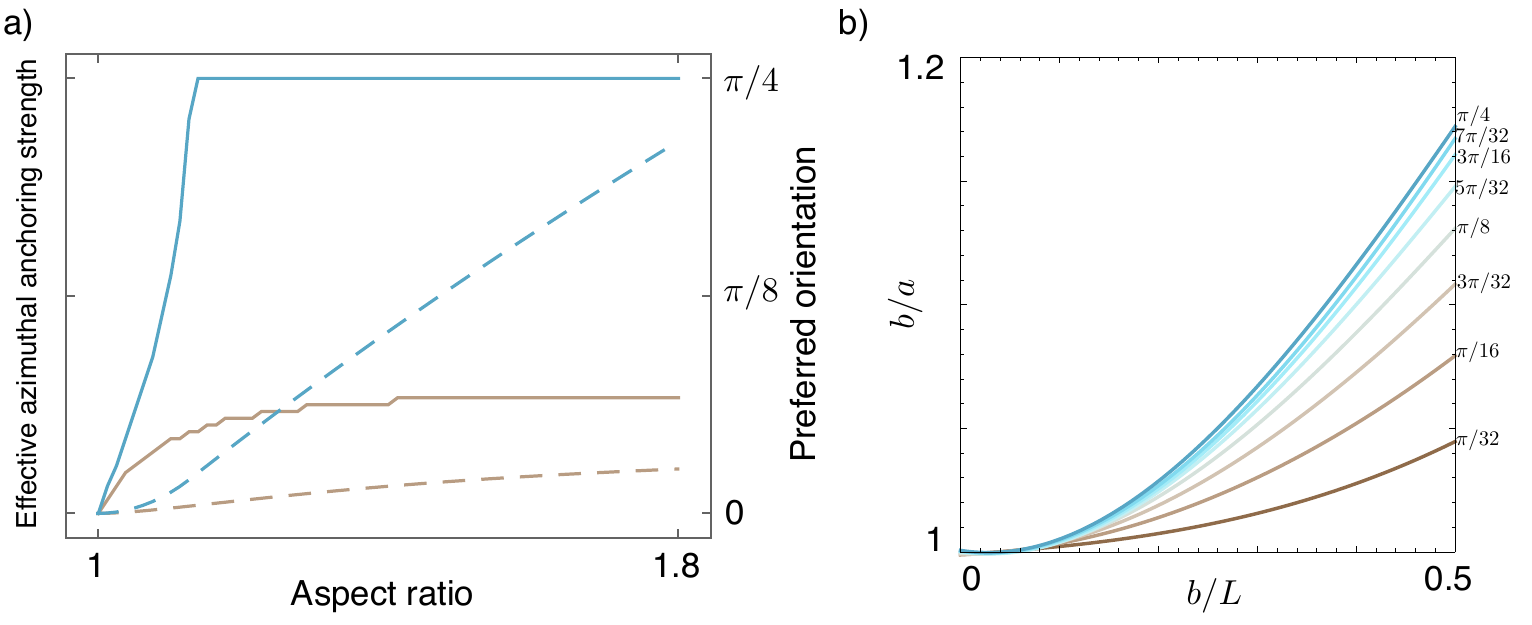}

\protect\caption{\label{fig:ellipsePhaseDiagram}(a) Magnitude (dashed, left axis)
and direction (solid, right axis) of azimuthal angle preference given
a growing aspect ratio for ellipse alignment angles of $\omega=\pi/4$
(cyan) and $\omega=\pi/8$ (brown) with $d/\lambda=1$, $L_{\theta}=0.01$,
and a coverage fraction equal to that of a circle with $R/\lambda=0.4$.
(b) Phase diagram of azimuthal angle preference of the entire unit
cell as a function of aspect ratio $b/a$ and semi major axis length
$b/\lambda$ with $\omega=\pi/4$, $d/\lambda=1$, and $L_{\theta}=0.01$.
Note these parameters do not maintain a constant coverage fraction.}
\end{figure*}

The preferred angle $\phi_{0}$ is therefore the result of a tension
between surface and bulk effects. We display the direction and strength
of the effective anchoring potential as a function of aspect ratio
in Fig. \ref{fig:ellipsePhaseDiagram}(a). Fig. \ref{fig:ellipsePhaseDiagram}(b)
presents a phase diagram for preferred angle as a function of the
semi-major axis and aspect ratio of the elliptical motifs while holding
ellipse rotation angle, cell depth, and anchoring potential constant.
As expected, increasing aspect ratio leads to alignment of the effective
easy axis with the ellipse long axis. Less intuitive is the fact that
smaller ellipses transition from alignment with the lattice to alignment
with the basis at lower aspect ratios. The decreased scale of the
bulk LC energy for small surface features drives this as unit cells
with small motifs have surface and bulk energies at the same scale.
Thus, increases in aspect ratio quickly shift the LC azimuthal preference
to align with the basis, overcoming the influence of the bulk.

\subsection{Numerical model\label{sub:Numerical-model}}

\begin{figure}
\includegraphics{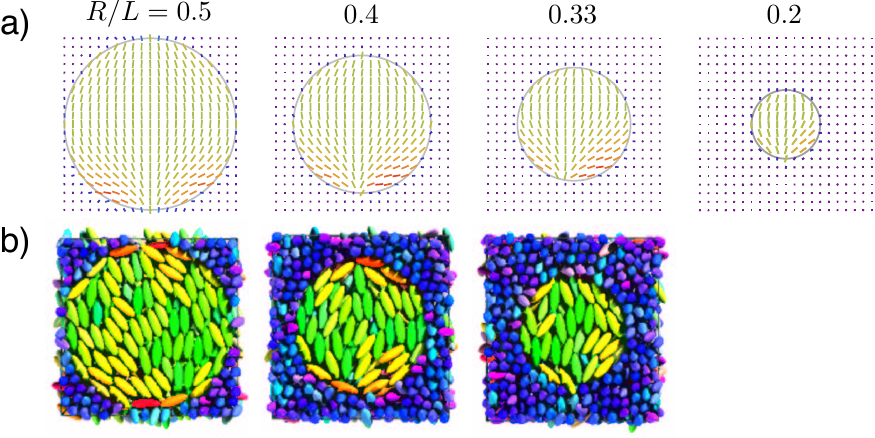}

\protect\caption{\label{fig:3Dtangent}(a) Results from 3D director minimization model.
Shown is the director orientation on the bottom face of the unit cell
for planar circle patterns of radius $R/\lambda=0.5$, $R/\lambda=0.4$,
$R/\lambda=0.33$, and $R/\lambda=0.2$. These simulations are performed
with $L_{\theta}=.001$, $d=1.0$, a final grid size of $20\times20\times21$,
and an initial guess that had the director aligned along the $y$-axis.
(b) Corresponding results from Monte Carlo simulations. }
\end{figure}

As briefly discussed earlier, the Monte Carlo simulations produced
configurations in which the particles tended to align tangentially
with the boundary of the circle pattern. These are shown in Fig. \ref{fig:3Dtangent}
and, together with the results of the previous sections, suggest that
the constant $\phi$ approximation may be too restrictive for this
system. We therefore performed a numerical minimization of the Frank
energy (\ref{eq:frank}) again using the two-constant approximation
but now allowing the director to vary arbitrarily in 3D. A Cartesian
representation of the director $\hat{n}=(n_{x},n_{y},n_{z})$ was
used and the unit length constraint $\hat{n}\cdot\hat{n}=1$ enforced
locally. The energy was discretized using second-order finite differences
and minimized using an adaptive gradient-descent relaxation method
with line searches. To improve convergence, successive refinement
was used: an initial guess is relaxed on a coarse grid, then interpolated
and relaxed onto successively finer grids. At each step, the system
was relaxed until the energy converged. 

Results from the relaxation model for a set of planar patterned surfaces
are shown in Fig. \ref{fig:3Dtangent}(a). An initial guess with the
director aligned along the $y$-axis was used. After relaxation, the
director adopts an orientation tangent to the pattern edges, as seen
in the corresponding Monte Carlo simulations, Fig. \ref{fig:3Dtangent}(b).
The tangential alignment arises because it corresponds to a twist
deformation across the vertical planar boundary, which is energetically
cheaper than a bend or splay deformation. The behavior breaks down
for smaller circle radii because the bend deformation required to
follow the arc of the feature becomes too energetically expensive.

\section{Conclusion\label{sec:Conclusions}}

This paper has considered the alignment behavior of square arrays
decorated with elliptical motifs, demonstrating that such patterns
can be used to create surfaces with controllable anchoring energy
and easy axis by varying the period of the pattern, ellipse orientation,
aspect ratio, and coverage fraction. Given the two-constant approximation
and the assumption of a constant azimuthal angle, the director configuration
and energy can be computed analytically. Depending on the circle radius,
cell depth, and polar anchoring strength of the alignment material,
the ground state may have azimuthal alignment along either of the
lattice vectors, or diagonally. 

Our study offers invaluable advice for applications because, while
the elliptical patterned surface offers a remarkable degree of control
over the anchoring properties, the design parameter space for the
pattern is large. Briefly, surfaces patterned by rotated ellipses
allow control of the azimuthal angle over a continuum of values between
the lattice vectors and the diagonal depending on the orientation
of the ellipse. In these cases, the ellipse aspect ratio controls
the effective anchoring energy. The behavior is nontrivial, however,
and includes regions of bistability, as well as an azimuthal anchoring
transition for some designs. 

Unlike for other patterned surfaces previously studied, the constant
azimuthal angle approximation is of only limited use. A numerical
study showed a preference for tangential alignment along the vertical-planar
boundary of the pattern in excellent agreement with Monte Carlo simulations. 

While the present study has considered a flat surface, the results
may also be important for experimentalists using arrays of posts to
align liquid crystals as has been reported in \citep{Kitson2002,Cavallaro19112013,C3SM53170H}.
It seems that only a very modest amount of anisotropy in the shape
of the pillars, for example using posts of elliptical cross section,
may break the square symmetry and lead to significantly better azimuthal
alignment if desired. 
\begin{acknowledgments}
\emph{The authors wish to thank Tufts University for a Tufts Collaborates!
seed grant that partly funded one of the co-authors (DBE). ADB was
part funded by a Graduate Student Summer Scholarship from Tufts University.
TJA is grateful to the Research Corporation for Science Advancement
for a Cottrell Scholar Award. }
\end{acknowledgments}

\appendix
\begin{widetext}

\section*{Appendix\label{sec:Boundary-condition-Fourier}}

To evaluate the Fourier coefficients for the easy axis, first substitute
$x\rightarrow x'+1/2$ and $y\rightarrow y'+1/2$, which allows (\ref{eq:surfcoef})
and (\ref{eq:domain}) to be rewritten as

\begin{equation}
S_{nm}=\frac{\pi}{2}\exp\big(\pi i(n+m)\big)\iint_{\mathcal{D}}\exp\big(2\pi i(nx'+my')\big)dx'dy',\label{eq:surfcoef1}
\end{equation}
and

\begin{equation}
\left(\frac{x'\cos\omega+y'\sin\omega}{a}\right)^{2}+\left(\frac{x'\sin\omega-y'\cos\omega}{b}\right)^{2}\le1,\label{eq:domain1}
\end{equation}
respectively. Note that the exponential pre-factor in (\ref{eq:surfcoef1})
is $(-1)^{n+m}$ since $n$ and $m$ are integers.

Next, rotate the coordinates via the transformations $(x'\cos\omega+y'\sin\omega)/a\rightarrow x''$
and $(x'\sin\omega-y'\cos\omega)/b\rightarrow y''$, and integrate
using these new coordinates,

\begin{equation}
\underset{^{(x'')^{2}+(y'')^{2}\le1}}{S_{nm}=\frac{ab\pi}{2}(-1)^{n+m}\iint\exp\Big(2\pi i\big((n\cos\omega+}m\sin\omega)ax''+(n\sin\omega-m\cos\omega)by''\big)\Big)dx''dy''.
\end{equation}
Define $a'=(n\cos\omega+m\sin\omega)a$, $b'=(n\sin\omega-m\cos\omega)b$,
$\alpha=(a',b')$, and $g(s)=\exp(is)$ and convert to polar coordinates
such that $(x'',y'')\rightarrow r\xi$ where $\xi=(\cos\theta,\sin\theta)$:

\begin{equation}
S_{nm}=\frac{ab\pi}{2}(-1)^{n+m}\int_{0}^{1}r\,dr\int_{\xi}g(2\pi r\alpha\cdot\xi)d\xi.
\end{equation}
Eqs. (1.2) and (1.5) from \citep{john2004} allow us to evaluate the
inner integral to find

\begin{equation}
S_{nm}=\frac{ab\pi}{2}(-1)^{n+m}\int_{0}^{1}2\pi rJ_{0}\left(2\pi r\sqrt{a'^{2}+b'^{2}}\right)dr.
\end{equation}
Finally, let $k=2\pi\sqrt{a'^{2}+b'^{2}}$ and $s=kr$, and note that
$\int xJ_{0}(x)dx=xJ_{1}(x)$ to obtain,

\begin{eqnarray}
S_{nm} & = & \frac{ab\pi}{2}(-1)^{n+m}\frac{2\pi}{k^{2}}\int_{0}^{k}sJ_{0}(s)ds\nonumber \\
 & = & \frac{ab\pi}{2}(-1)^{n+m}\frac{2\pi}{k}J_{1}(k)\nonumber \\
 & = & \frac{ab\pi}{2}(-1)^{n+m}\frac{J_{1}\left(2\pi\sqrt{a'^{2}+b'^{2}}\right)}{\sqrt{a'^{2}+b'^{2}}},
\end{eqnarray}
the result stated in Eq. (\ref{eq:surfcoefsol}).

\end{widetext}


\begin{thebibliography}{40}
\expandafter\ifx\csname natexlab\endcsname\relax\def\natexlab#1{#1}\fi
\expandafter\ifx\csname bibnamefont\endcsname\relax
  \def\bibnamefont#1{#1}\fi
\expandafter\ifx\csname bibfnamefont\endcsname\relax
  \def\bibfnamefont#1{#1}\fi
\expandafter\ifx\csname citenamefont\endcsname\relax
  \def\citenamefont#1{#1}\fi
\expandafter\ifx\csname url\endcsname\relax
  \def\url#1{\texttt{#1}}\fi
\expandafter\ifx\csname urlprefix\endcsname\relax\def\urlprefix{URL }\fi
\providecommand{\bibinfo}[2]{#2}
\providecommand{\eprint}[2][]{\url{#2}}

\bibitem[{\citenamefont{Jerome}(1991)}]{Jerome1991}
\bibinfo{author}{\bibfnamefont{B.}~\bibnamefont{Jerome}},
  \bibinfo{journal}{Rep. Prog. Phys.} \textbf{\bibinfo{volume}{54}},
  \bibinfo{pages}{391} (\bibinfo{year}{1991}).

\bibitem[{\citenamefont{Spencer and Care}(2006)}]{spencer2006}
\bibinfo{author}{\bibfnamefont{T.~J.} \bibnamefont{Spencer}} \bibnamefont{and}
  \bibinfo{author}{\bibfnamefont{C.~M.} \bibnamefont{Care}},
  \bibinfo{journal}{Phys. Rev. E} \textbf{\bibinfo{volume}{74}},
  \bibinfo{pages}{061708} (\bibinfo{year}{2006}).

\bibitem[{\citenamefont{Kondrat et~al.}(2003)\citenamefont{Kondrat,
  Poniewierski, and Harnau}}]{kondrat2003}
\bibinfo{author}{\bibfnamefont{S.}~\bibnamefont{Kondrat}},
  \bibinfo{author}{\bibfnamefont{A.}~\bibnamefont{Poniewierski}},
  \bibnamefont{and} \bibinfo{author}{\bibfnamefont{L.}~\bibnamefont{Harnau}},
  \bibinfo{journal}{The European Physical Journal E}
  \textbf{\bibinfo{volume}{10}}, \bibinfo{pages}{163} (\bibinfo{year}{2003}).

\bibitem[{\citenamefont{Barbero et~al.}(1992)\citenamefont{Barbero, Beica,
  Alexe-Ionescu, and Moldovan}}]{barbero1992}
\bibinfo{author}{\bibfnamefont{G.}~\bibnamefont{Barbero}},
  \bibinfo{author}{\bibfnamefont{T.}~\bibnamefont{Beica}},
  \bibinfo{author}{\bibfnamefont{A.~L.} \bibnamefont{Alexe-Ionescu}},
  \bibnamefont{and} \bibinfo{author}{\bibfnamefont{R.}~\bibnamefont{Moldovan}},
  \bibinfo{journal}{Journal de Physique II} \textbf{\bibinfo{volume}{2}},
  \bibinfo{pages}{2011} (\bibinfo{year}{1992}).

\bibitem[{\citenamefont{Anquetil-Deck et~al.}(2012)\citenamefont{Anquetil-Deck,
  Cleaver, and Atherton}}]{Anquetil-Deck2012}
\bibinfo{author}{\bibfnamefont{C.}~\bibnamefont{Anquetil-Deck}},
  \bibinfo{author}{\bibfnamefont{D.~J.} \bibnamefont{Cleaver}},
  \bibnamefont{and} \bibinfo{author}{\bibfnamefont{T.~J.}
  \bibnamefont{Atherton}}, \bibinfo{journal}{Phys. Rev. E}
  \textbf{\bibinfo{volume}{86}}, \bibinfo{pages}{041707}
  (\bibinfo{year}{2012}).

\bibitem[{\citenamefont{Kim et~al.}(2001)\citenamefont{Kim, Yoneya, Yamamoto,
  and Yokoyama}}]{Kim2001}
\bibinfo{author}{\bibfnamefont{J.~H.} \bibnamefont{Kim}},
  \bibinfo{author}{\bibfnamefont{M.}~\bibnamefont{Yoneya}},
  \bibinfo{author}{\bibfnamefont{J.}~\bibnamefont{Yamamoto}}, \bibnamefont{and}
  \bibinfo{author}{\bibfnamefont{H.}~\bibnamefont{Yokoyama}},
  \bibinfo{journal}{Appl. Phys. Lett.} \textbf{\bibinfo{volume}{78}},
  \bibinfo{pages}{3055} (\bibinfo{year}{2001}).

\bibitem[{\citenamefont{Yi et~al.}(2008)\citenamefont{Yi, Khire, Bowman,
  MacLennan, and Clark}}]{Yi2008}
\bibinfo{author}{\bibfnamefont{Y.~W.} \bibnamefont{Yi}},
  \bibinfo{author}{\bibfnamefont{V.}~\bibnamefont{Khire}},
  \bibinfo{author}{\bibfnamefont{C.~N.} \bibnamefont{Bowman}},
  \bibinfo{author}{\bibfnamefont{J.~E.} \bibnamefont{MacLennan}},
  \bibnamefont{and} \bibinfo{author}{\bibfnamefont{N.~A.} \bibnamefont{Clark}},
  \bibinfo{journal}{J. Appl. Phys.} \textbf{\bibinfo{volume}{103}},
  \bibinfo{pages}{1} (\bibinfo{year}{2008}).

\bibitem[{\citenamefont{Yoneya et~al.}(2002)\citenamefont{Yoneya, Kim, and
  Yokoyama}}]{Yoneya2002a}
\bibinfo{author}{\bibfnamefont{M.}~\bibnamefont{Yoneya}},
  \bibinfo{author}{\bibfnamefont{J.~H.} \bibnamefont{Kim}}, \bibnamefont{and}
  \bibinfo{author}{\bibfnamefont{H.}~\bibnamefont{Yokoyama}},
  \bibinfo{journal}{Appl. Phys. Lett.} \textbf{\bibinfo{volume}{80}},
  \bibinfo{pages}{374} (\bibinfo{year}{2002}).

\bibitem[{\citenamefont{Lee and Clark}(2001)}]{Lee2001}
\bibinfo{author}{\bibfnamefont{B.}~\bibnamefont{Lee}} \bibnamefont{and}
  \bibinfo{author}{\bibfnamefont{N.~A.} \bibnamefont{Clark}},
  \bibinfo{journal}{Science} \textbf{\bibinfo{volume}{291}},
  \bibinfo{pages}{2576} (\bibinfo{year}{2001}).

\bibitem[{\citenamefont{Kim et~al.}(2002)\citenamefont{Kim, Yoneya, and
  Yokoyama}}]{Kim2002}
\bibinfo{author}{\bibfnamefont{J.-H.} \bibnamefont{Kim}},
  \bibinfo{author}{\bibfnamefont{M.}~\bibnamefont{Yoneya}}, \bibnamefont{and}
  \bibinfo{author}{\bibfnamefont{H.}~\bibnamefont{Yokoyama}},
  \bibinfo{journal}{Nature} \textbf{\bibinfo{volume}{420}},
  \bibinfo{pages}{159} (\bibinfo{year}{2002}).

\bibitem[{\citenamefont{Stalder and Schadt}(2003)}]{Stalder2003}
\bibinfo{author}{\bibfnamefont{M.}~\bibnamefont{Stalder}} \bibnamefont{and}
  \bibinfo{author}{\bibfnamefont{M.}~\bibnamefont{Schadt}},
  \bibinfo{journal}{Liq. Cryst.} \textbf{\bibinfo{volume}{30}},
  \bibinfo{pages}{285} (\bibinfo{year}{2003}).

\bibitem[{\citenamefont{Bryan-Brown}(2000)}]{Bryan-brown2000}
\bibinfo{author}{\bibfnamefont{G.~P.} \bibnamefont{Bryan-Brown}},
  \bibinfo{journal}{Displays} \textbf{\bibinfo{volume}{27}},
  \bibinfo{pages}{37} (\bibinfo{year}{2000}).

\bibitem[{\citenamefont{Kitson and Geisow}(2002)}]{Kitson2002}
\bibinfo{author}{\bibfnamefont{S.}~\bibnamefont{Kitson}} \bibnamefont{and}
  \bibinfo{author}{\bibfnamefont{A.}~\bibnamefont{Geisow}},
  \bibinfo{journal}{Appl. Phys. Lett.} \textbf{\bibinfo{volume}{80}},
  \bibinfo{pages}{3635} (\bibinfo{year}{2002}).

\bibitem[{\citenamefont{Hwang and Rey}(2006)}]{Hwang2006}
\bibinfo{author}{\bibfnamefont{D.~K.} \bibnamefont{Hwang}} \bibnamefont{and}
  \bibinfo{author}{\bibfnamefont{A.~D.} \bibnamefont{Rey}},
  \bibinfo{journal}{Soc. Ind. Appl. Math.} \textbf{\bibinfo{volume}{67}},
  \bibinfo{pages}{214} (\bibinfo{year}{2006}).

\bibitem[{\citenamefont{Lowe et~al.}(2010)\citenamefont{Lowe, Ozer, Bai,
  Bertics, and Abbott}}]{Lowe2010}
\bibinfo{author}{\bibfnamefont{A.~M.} \bibnamefont{Lowe}},
  \bibinfo{author}{\bibfnamefont{B.~H.} \bibnamefont{Ozer}},
  \bibinfo{author}{\bibfnamefont{Y.}~\bibnamefont{Bai}},
  \bibinfo{author}{\bibfnamefont{P.~J.} \bibnamefont{Bertics}},
  \bibnamefont{and} \bibinfo{author}{\bibfnamefont{N.~L.}
  \bibnamefont{Abbott}}, \bibinfo{journal}{ACS Appl. Mater. Interfaces}
  \textbf{\bibinfo{volume}{2}}, \bibinfo{pages}{722} (\bibinfo{year}{2010}).

\bibitem[{\citenamefont{Ruan et~al.}(2003)\citenamefont{Ruan, Sambles, and
  Stewart}}]{Ruan2003}
\bibinfo{author}{\bibfnamefont{L.~Z.} \bibnamefont{Ruan}},
  \bibinfo{author}{\bibfnamefont{J.~R.} \bibnamefont{Sambles}},
  \bibnamefont{and} \bibinfo{author}{\bibfnamefont{I.~W.}
  \bibnamefont{Stewart}}, \bibinfo{journal}{Phys. Rev. Lett.}
  \textbf{\bibinfo{volume}{91}}, \bibinfo{pages}{033901}
  (\bibinfo{year}{2003}).

\bibitem[{\citenamefont{Wei et~al.}(2009)\citenamefont{Wei, Weirich,
  Alkeskjold, and Bjarklev}}]{Wei2009}
\bibinfo{author}{\bibfnamefont{L.}~\bibnamefont{Wei}},
  \bibinfo{author}{\bibfnamefont{J.}~\bibnamefont{Weirich}},
  \bibinfo{author}{\bibfnamefont{T.~T.} \bibnamefont{Alkeskjold}},
  \bibnamefont{and} \bibinfo{author}{\bibfnamefont{A.}~\bibnamefont{Bjarklev}},
  \bibinfo{journal}{Opt. Lett.} \textbf{\bibinfo{volume}{34}},
  \bibinfo{pages}{3818} (\bibinfo{year}{2009}).

\bibitem[{\citenamefont{Bechtold and Oliveira}(2005)}]{Bechtold2005}
\bibinfo{author}{\bibfnamefont{I.~H.} \bibnamefont{Bechtold}} \bibnamefont{and}
  \bibinfo{author}{\bibfnamefont{E.~A.} \bibnamefont{Oliveira}},
  \bibinfo{journal}{Mol. Cryst. Liq. Cryst.} \textbf{\bibinfo{volume}{442}},
  \bibinfo{pages}{41} (\bibinfo{year}{2005}).

\bibitem[{\citenamefont{Varghese et~al.}(2004)\citenamefont{Varghese,
  Narayanankutty, Bastiaansen, Crawford, and Broer}}]{Varghese2004}
\bibinfo{author}{\bibfnamefont{S.}~\bibnamefont{Varghese}},
  \bibinfo{author}{\bibfnamefont{S.}~\bibnamefont{Narayanankutty}},
  \bibinfo{author}{\bibfnamefont{C.~W.~M.} \bibnamefont{Bastiaansen}},
  \bibinfo{author}{\bibfnamefont{G.~P.} \bibnamefont{Crawford}},
  \bibnamefont{and} \bibinfo{author}{\bibfnamefont{D.~J.} \bibnamefont{Broer}},
  \bibinfo{journal}{Advanced Materials} \textbf{\bibinfo{volume}{16}},
  \bibinfo{pages}{1600} (\bibinfo{year}{2004}).

\bibitem[{\citenamefont{Schadt et~al.}(1992)\citenamefont{Schadt, Schmitt,
  Kozinkov, and Chigrinov}}]{Schadt1992}
\bibinfo{author}{\bibfnamefont{M.}~\bibnamefont{Schadt}},
  \bibinfo{author}{\bibfnamefont{K.}~\bibnamefont{Schmitt}},
  \bibinfo{author}{\bibfnamefont{V.}~\bibnamefont{Kozinkov}}, \bibnamefont{and}
  \bibinfo{author}{\bibfnamefont{V.}~\bibnamefont{Chigrinov}},
  \bibinfo{journal}{Japanese Journal of Applied Physics}
  \textbf{\bibinfo{volume}{31}}, \bibinfo{pages}{2155} (\bibinfo{year}{1992}).

\bibitem[{\citenamefont{Lee et~al.}(2004)\citenamefont{Lee, Zhang, Sheng, Kwok,
  and Tsui}}]{Lee2004}
\bibinfo{author}{\bibfnamefont{F.~K.} \bibnamefont{Lee}},
  \bibinfo{author}{\bibfnamefont{B.}~\bibnamefont{Zhang}},
  \bibinfo{author}{\bibfnamefont{P.}~\bibnamefont{Sheng}},
  \bibinfo{author}{\bibfnamefont{H.~S.} \bibnamefont{Kwok}}, \bibnamefont{and}
  \bibinfo{author}{\bibfnamefont{O.~K.~C.} \bibnamefont{Tsui}},
  \bibinfo{journal}{Appl. Phys. Lett.} \textbf{\bibinfo{volume}{85}},
  \bibinfo{pages}{5556} (\bibinfo{year}{2004}).

\bibitem[{\citenamefont{Gupta and Abbott}(1997)}]{Gupta1997}
\bibinfo{author}{\bibfnamefont{V.~K.} \bibnamefont{Gupta}} \bibnamefont{and}
  \bibinfo{author}{\bibfnamefont{N.~L.} \bibnamefont{Abbott}},
  \bibinfo{journal}{Science} \textbf{\bibinfo{volume}{276}},
  \bibinfo{pages}{1533} (\bibinfo{year}{1997}).

\bibitem[{\citenamefont{Cheng et~al.}(2000)\citenamefont{Cheng, Batchelder,
  Evans, Henderson, Lydon, and Ogier}}]{Cheng2000}
\bibinfo{author}{\bibfnamefont{Y.~L.} \bibnamefont{Cheng}},
  \bibinfo{author}{\bibfnamefont{D.~N.} \bibnamefont{Batchelder}},
  \bibinfo{author}{\bibfnamefont{S.~D.} \bibnamefont{Evans}},
  \bibinfo{author}{\bibfnamefont{J.~R.} \bibnamefont{Henderson}},
  \bibinfo{author}{\bibfnamefont{J.~E.} \bibnamefont{Lydon}}, \bibnamefont{and}
  \bibinfo{author}{\bibfnamefont{S.~D.} \bibnamefont{Ogier}},
  \bibinfo{journal}{Liq. Cryst.} \textbf{\bibinfo{volume}{27}},
  \bibinfo{pages}{1267} (\bibinfo{year}{2000}).

\bibitem[{\citenamefont{Prompinit et~al.}(2010)\citenamefont{Prompinit,
  Achalkumar, Bramble, Bushby, W\"{a}lti, and Evans}}]{Prompinit2010}
\bibinfo{author}{\bibfnamefont{P.}~\bibnamefont{Prompinit}},
  \bibinfo{author}{\bibfnamefont{A.~S.} \bibnamefont{Achalkumar}},
  \bibinfo{author}{\bibfnamefont{J.~P.} \bibnamefont{Bramble}},
  \bibinfo{author}{\bibfnamefont{R.~J.} \bibnamefont{Bushby}},
  \bibinfo{author}{\bibfnamefont{C.}~\bibnamefont{W\"{a}lti}},
  \bibnamefont{and} \bibinfo{author}{\bibfnamefont{S.~D.} \bibnamefont{Evans}},
  \bibinfo{journal}{ACS Appl. Mater. Interfaces} \textbf{\bibinfo{volume}{2}},
  \bibinfo{pages}{3686} (\bibinfo{year}{2010}).

\bibitem[{\citenamefont{Zheng et~al.}(2013)\citenamefont{Zheng, Chiang, and
  Underwood}}]{Zheng2013}
\bibinfo{author}{\bibfnamefont{W.}~\bibnamefont{Zheng}},
  \bibinfo{author}{\bibfnamefont{C.-Y.} \bibnamefont{Chiang}},
  \bibnamefont{and}
  \bibinfo{author}{\bibfnamefont{I.}~\bibnamefont{Underwood}},
  \bibinfo{journal}{Thin Solid Films} \textbf{\bibinfo{volume}{545}},
  \bibinfo{pages}{371} (\bibinfo{year}{2013}).

\bibitem[{\citenamefont{Liu et~al.}(2012)\citenamefont{Liu, Bastiaansen, den
  Toonder, and Broer}}]{Liu2012}
\bibinfo{author}{\bibfnamefont{D.}~\bibnamefont{Liu}},
  \bibinfo{author}{\bibfnamefont{C.~W.~M.} \bibnamefont{Bastiaansen}},
  \bibinfo{author}{\bibfnamefont{J.~M.~J.} \bibnamefont{den Toonder}},
  \bibnamefont{and} \bibinfo{author}{\bibfnamefont{D.~J.} \bibnamefont{Broer}},
  \bibinfo{journal}{Macromolecules} \textbf{\bibinfo{volume}{45}},
  \bibinfo{pages}{8005} (\bibinfo{year}{2012}).

\bibitem[{\citenamefont{Park and Park}(2008)}]{Park2008}
\bibinfo{author}{\bibfnamefont{M.~J.} \bibnamefont{Park}} \bibnamefont{and}
  \bibinfo{author}{\bibfnamefont{O.~O.} \bibnamefont{Park}},
  \bibinfo{journal}{Microelectron. Eng.} \textbf{\bibinfo{volume}{85}},
  \bibinfo{pages}{2261} (\bibinfo{year}{2008}).

\bibitem[{\citenamefont{Wilderbeek et~al.}(2003)\citenamefont{Wilderbeek,
  Teunissen, Bastiaansen, and Broer}}]{Wilderbeek2003}
\bibinfo{author}{\bibfnamefont{H.~T.~A.} \bibnamefont{Wilderbeek}},
  \bibinfo{author}{\bibfnamefont{J.~P.} \bibnamefont{Teunissen}},
  \bibinfo{author}{\bibfnamefont{C.~W.~M.} \bibnamefont{Bastiaansen}},
  \bibnamefont{and} \bibinfo{author}{\bibfnamefont{D.~J.} \bibnamefont{Broer}},
  \bibinfo{journal}{Advanced Materials} \textbf{\bibinfo{volume}{15}},
  \bibinfo{pages}{985} (\bibinfo{year}{2003}).

\bibitem[{\citenamefont{Atherton and Sambles}(2006)}]{Atherton2006}
\bibinfo{author}{\bibfnamefont{T.~J.} \bibnamefont{Atherton}} \bibnamefont{and}
  \bibinfo{author}{\bibfnamefont{J.}~\bibnamefont{Sambles}},
  \bibinfo{journal}{Phys. Rev. E} \textbf{\bibinfo{volume}{74}},
  \bibinfo{pages}{022701} (\bibinfo{year}{2006}).

\bibitem[{\citenamefont{Atherton et~al.}(2009)\citenamefont{Atherton, Sambles,
  Bramble, Henderson, and Evans}}]{Atherton2009}
\bibinfo{author}{\bibfnamefont{T.~J.} \bibnamefont{Atherton}},
  \bibinfo{author}{\bibfnamefont{J.~R.} \bibnamefont{Sambles}},
  \bibinfo{author}{\bibfnamefont{J.~P.} \bibnamefont{Bramble}},
  \bibinfo{author}{\bibfnamefont{J.~R.} \bibnamefont{Henderson}},
  \bibnamefont{and} \bibinfo{author}{\bibfnamefont{S.~D.} \bibnamefont{Evans}},
  \bibinfo{journal}{Liq. Cryst.} \textbf{\bibinfo{volume}{36}},
  \bibinfo{pages}{353} (\bibinfo{year}{2009}).

\bibitem[{\citenamefont{Atherton}(2010)}]{Atherton2010}
\bibinfo{author}{\bibfnamefont{T.}~\bibnamefont{Atherton}},
  \bibinfo{journal}{Liq. Cryst.} \textbf{\bibinfo{volume}{37}},
  \bibinfo{pages}{1225} (\bibinfo{year}{2010}).

\bibitem[{\citenamefont{Ledney and Tarnavskyy}(2011)}]{Ledney2011}
\bibinfo{author}{\bibfnamefont{M.}~\bibnamefont{Ledney}} \bibnamefont{and}
  \bibinfo{author}{\bibfnamefont{O.}~\bibnamefont{Tarnavskyy}},
  \bibinfo{journal}{Soft Matter} \textbf{\bibinfo{volume}{56}},
  \bibinfo{pages}{880} (\bibinfo{year}{2011}).

\bibitem[{\citenamefont{Kondrat et~al.}(2005)\citenamefont{Kondrat,
  Poniewierski, and Harnau}}]{Kondrat2005}
\bibinfo{author}{\bibfnamefont{S.}~\bibnamefont{Kondrat}},
  \bibinfo{author}{\bibfnamefont{A.}~\bibnamefont{Poniewierski}},
  \bibnamefont{and} \bibinfo{author}{\bibfnamefont{L.}~\bibnamefont{Harnau}},
  \bibinfo{journal}{Liq. Cryst.} \textbf{\bibinfo{volume}{32}},
  \bibinfo{pages}{95} (\bibinfo{year}{2005}).

\bibitem[{\citenamefont{Yi and Clark}(2013)}]{Yi2013}
\bibinfo{author}{\bibfnamefont{Y.}~\bibnamefont{Yi}} \bibnamefont{and}
  \bibinfo{author}{\bibfnamefont{N.~A.} \bibnamefont{Clark}},
  \bibinfo{journal}{Liq. Cryst.} \textbf{\bibinfo{volume}{40}},
  \bibinfo{pages}{1736} (\bibinfo{year}{2013}).

\bibitem[{\citenamefont{Anquetil-Deck et~al.}(2013)\citenamefont{Anquetil-Deck,
  Cleaver, Bramble, and Atherton}}]{Anquetil-Deck2013}
\bibinfo{author}{\bibfnamefont{C.}~\bibnamefont{Anquetil-Deck}},
  \bibinfo{author}{\bibfnamefont{D.~J.} \bibnamefont{Cleaver}},
  \bibinfo{author}{\bibfnamefont{J.~P.} \bibnamefont{Bramble}},
  \bibnamefont{and} \bibinfo{author}{\bibfnamefont{T.~J.}
  \bibnamefont{Atherton}}, \bibinfo{journal}{Phys. Rev. E}
  \textbf{\bibinfo{volume}{88}}, \bibinfo{pages}{012501}
  (\bibinfo{year}{2013}).

\bibitem[{\citenamefont{Harnau et~al.}(2007)\citenamefont{Harnau, Kondrat, and
  Poniewierski}}]{Harnau2007}
\bibinfo{author}{\bibfnamefont{L.}~\bibnamefont{Harnau}},
  \bibinfo{author}{\bibfnamefont{S.}~\bibnamefont{Kondrat}}, \bibnamefont{and}
  \bibinfo{author}{\bibfnamefont{A.}~\bibnamefont{Poniewierski}},
  \bibinfo{journal}{Phys. Rev. E} \textbf{\bibinfo{volume}{76}},
  \bibinfo{pages}{051701} (\bibinfo{year}{2007}).

\bibitem[{\citenamefont{Ferjani et~al.}(2010)\citenamefont{Ferjani, Choi,
  Pendery, Petschek, and Rosenblatt}}]{Ferjani2010}
\bibinfo{author}{\bibfnamefont{S.}~\bibnamefont{Ferjani}},
  \bibinfo{author}{\bibfnamefont{Y.}~\bibnamefont{Choi}},
  \bibinfo{author}{\bibfnamefont{J.}~\bibnamefont{Pendery}},
  \bibinfo{author}{\bibfnamefont{R.}~\bibnamefont{Petschek}}, \bibnamefont{and}
  \bibinfo{author}{\bibfnamefont{C.}~\bibnamefont{Rosenblatt}},
  \bibinfo{journal}{Phys. Rev. Lett.} \textbf{\bibinfo{volume}{104}}
  (\bibinfo{year}{2010}).

\bibitem[{\citenamefont{Cavallaro et~al.}(2013)\citenamefont{Cavallaro, Gharbi,
  Beller, {\v C}opar, Shi, Baumgart, Yang, Kamien, and
  Stebe}}]{Cavallaro19112013}
\bibinfo{author}{\bibfnamefont{M.}~\bibnamefont{Cavallaro}},
  \bibinfo{author}{\bibfnamefont{M.~A.} \bibnamefont{Gharbi}},
  \bibinfo{author}{\bibfnamefont{D.~A.} \bibnamefont{Beller}},
  \bibinfo{author}{\bibfnamefont{S.}~\bibnamefont{{\v C}opar}},
  \bibinfo{author}{\bibfnamefont{Z.}~\bibnamefont{Shi}},
  \bibinfo{author}{\bibfnamefont{T.}~\bibnamefont{Baumgart}},
  \bibinfo{author}{\bibfnamefont{S.}~\bibnamefont{Yang}},
  \bibinfo{author}{\bibfnamefont{R.~D.} \bibnamefont{Kamien}},
  \bibnamefont{and} \bibinfo{author}{\bibfnamefont{K.~J.} \bibnamefont{Stebe}},
  \bibinfo{journal}{PNAS} \textbf{\bibinfo{volume}{110}},
  \bibinfo{pages}{18804} (\bibinfo{year}{2013}).

\bibitem[{\citenamefont{Lohr et~al.}(2014)\citenamefont{Lohr, Cavallaro,
  Beller, Stebe, Kamien, Collings, and Yodh}}]{C3SM53170H}
\bibinfo{author}{\bibfnamefont{M.~A.} \bibnamefont{Lohr}},
  \bibinfo{author}{\bibfnamefont{M.}~\bibnamefont{Cavallaro}},
  \bibinfo{author}{\bibfnamefont{D.~A.} \bibnamefont{Beller}},
  \bibinfo{author}{\bibfnamefont{K.~J.} \bibnamefont{Stebe}},
  \bibinfo{author}{\bibfnamefont{R.~D.} \bibnamefont{Kamien}},
  \bibinfo{author}{\bibfnamefont{P.~J.} \bibnamefont{Collings}},
  \bibnamefont{and} \bibinfo{author}{\bibfnamefont{A.~G.} \bibnamefont{Yodh}},
  \bibinfo{journal}{Soft Matter} \textbf{\bibinfo{volume}{10}},
  \bibinfo{pages}{3477} (\bibinfo{year}{2014}).

\bibitem[{\citenamefont{John}(2004)}]{john2004}
\bibinfo{author}{\bibfnamefont{F.}~\bibnamefont{John}},
  \emph{\bibinfo{title}{Plane waves and spherical means applied to partial
  differential equations}} (\bibinfo{publisher}{Dover Publications},
  \bibinfo{year}{2004}).

\end{thebibliography}
\end{document}